\documentclass[a4paper,twoside,12pt]{article}
\input{obsjkt.sty}

\newcommand{\Msun}{\ensuremath{~{\rm M}_\odot}}                   
\newcommand{\Rsun}{\ensuremath{~{\rm R}_\odot}}                   
\newcommand{\rhosun}{\ensuremath{~\rho_\odot}}                    
\newcommand{\EBV}{\ensuremath{E(B\!-\!V)}}                        
\newcommand{\Grp}{\ensuremath{G_{\rm RP}}}                        
\newcommand{\degr}{\ensuremath{^\circ}}                           
\renewcommand{\kms}{~km~s$^{-1}$}                                 
\renewcommand{\cd}{~d$^{-1}$}                                     
\newcommand{\chir}{\ensuremath{\chi_\nu^{\,2}}}                   
\newcommand{\gaia}{\textit{Gaia}}                                 
\newcommand{\targ}{EY~Cep}
\newcommand{\targfull}{EY~Cephei}

\newcommand{\Msunnom}{\hbox{$\mathcal{M}^{\rm N}_\odot$}}
\newcommand{\Rsunnom}{\hbox{$\mathcal{R}^{\rm N}_\odot$}}
\newcommand{\Lsunnom}{\hbox{$\mathcal{L}^{\rm N}_\odot$}}

\usepackage{rotating}

\begin{document} 

\OBSheader{Rediscussion of eclipsing binaries: \targ}{J.\ Southworth}{2026 August}

\OBStitle{Rediscussion of eclipsing binaries. Paper 32. \\ The eccentric F-type system EY Cephei}

\OBSauth{John Southworth}

\OBSinstone{Astrophysics Group, Keele University, Staffordshire, ST5 5BG, UK}


\OBSabstract{\targ\ is a detached eclipsing binary containing two F0~V stars in an orbit with a period of 7.97~d and an eccentricity of 0.440. We determine the physical properties of the system using light-curves from the Transiting Exoplanet Survey Satellite and published spectroscopic measurements. We find masses of $1.523 \pm 0.008$\Msun\ and $1.494 \pm 0.014$\Msun, radii of $1.491 \pm 0.004$\Rsun\ and $1.446 \pm 0.004$\Rsun, and temperatures of $7070 \pm 170$~K and $6990 \pm 150$~K. We calculate the system's distance to be $300.3 \pm 3.8$~pc, in excellent agreement with the \gaia\ DR3 parallax, and estimate its age to be 220~Myr. We find no evidence for pulsations, but there are hints of variation in the eclipse times potentially attributable to a third body on a longer-period eccentric orbit.}


\section*{Introduction}

Eclipsing binary star systems (EBs) are of fundamental importance to astrophysics as a direct source of mass and radius measurements of normal stars. Detached eclipsing binaries (dEBs) are particularly useful as their components have evolved as single stars so can be used to check and calibrate theoretical models of single-star evolution \cite{Andersen91aarv,Torres++10aarv}. The current work is part of a series of papers \cite{Me20obs} in which primarily space-based light-curves \cite{Me21univ} are used to refine the properties of known dEBs. The properties of these objects are collected in the Detached Eclipsing Binary Catalogue \cite{Me15aspc} (DEBCat\footnote{\texttt{https://www.astro.keele.ac.uk/jkt/debcat/}}).

In this work we present an analysis of \targfull\ (see Table~\ref{tab:info}), which contains two F0~V stars on an eccentric orbit. The target was originally chosen as possibly containing pulsating stars, although this was not confirmed. We summarise the observational history of the system, present the new observational material, analyse the available light and radial velocity (RV) curves, and determine the physical properties and age of the component stars.


\section*{\targfull}


Eclipses in the \targ\ system were reported in 1958 by Strohmeier \cite{Strohmeier58kvrsb}. Stroh-meier \cite{Strohmeier63ibvs} gave an orbital period of 5.51672~d, which has turned out to be incorrect. Lacy \cite{Lacy85ibvs,Lacy90ibvs} reported the discovery of double lines in the spectrum of \targ, with the lines from the two stars having approximately equal widths and depths. Lacy \cite{Lacy92aj} measured its $V$ magnitude, and $B-V$ and $U-B$ colour indices outside eclipse. Lacy \cite{Lacy02aj} provided magnitudes and colour indices in the $uvby\beta$ system.

Lacy et al.\ \cite{Lacy+06aj} (hereafter L06) presented the only detailed analysis of \targ\ published so far, including a revision and confirmation of the orbital period of 7.971~d. Their spectrographic material comprised five spectra from the 2.1~m telescope and coud\'e spectrometer at the Kitt Peak National Observatory (KPNO), and 47 from the 1.5~m Tillinghast reflector, \'echelle spectrograph and Reticon detector at the Fred Lawrence Whipple Observatory (FLWO). Each yielded RV measurements for both stars. The projected rotational velocities were found to be $10 \pm 1$\kms\ for both stars. From the $UBV$ and $uvby\beta$ photometry L86 deduced spectral types of F0~V for both components, a temperature of $T_{\rm eff,A} = 7090 \pm 150$~K for star~A (the star eclipsed at phase zero), and a temperature of $T_{\rm eff,B} = 6970 \pm 150$~K for star~B (its companion). 

\begin{table}[t]
\caption{\em Basic information on \targfull. 
The $BV$ magnitudes are each the mean of 115 individual measurements \cite{Hog+00aa} distributed approximately 
randomly in orbital phase. The $JHK_s$ magnitudes from 2MASS \cite{Cutri+03book} were obtained at an orbital 
phase of 0.947, which is close to but confidently before the beginning of primary eclipse. \label{tab:info}}
\centering
\begin{tabular}{lll}
{\em Property}                            & {\em Value}                 & {\em Reference}                      \\[3pt]
Right ascension (J2000)                   & 03 40 04.07                 & \citenum{Gaia23aa}                   \\
Declination (J2000)                       & $+$81 01 09.0               & \citenum{Gaia23aa}                   \\
\textit{Gaia} DR3 designation             & 568183560050534016          & \citenum{Gaia21aa}                   \\
\textit{Gaia} DR3 parallax (mas)          & $3.3202 \pm 0.0135$         & \citenum{Gaia21aa}                   \\          
TESS\ Input Catalog designation           & TIC 297899564               & \citenum{Stassun+19aj}               \\
$U$ magnitude                             & $10.126\pm 0.010$           & \citenum{Lacy92aj}                   \\          
$B$ magnitude                             & $10.162\pm 0.009$           & \citenum{Lacy92aj}                   \\          
$V$ magnitude                             & $9.795 \pm 0.007$           & \citenum{Lacy92aj}                   \\          
$J$ magnitude                             & $9.035 \pm 0.030$           & \citenum{Cutri+03book}               \\
$H$ magnitude                             & $8.977 \pm 0.031$           & \citenum{Cutri+03book}               \\
$K_s$ magnitude                           & $8.841 \pm 0.020$           & \citenum{Cutri+03book}               \\
Spectral type                             & F0~V + F0~V                 & \citenum{Lacy+06aj}                  \\[3pt]       
\end{tabular}
\end{table}


L06 obtained light-curves comprising 6907 measurements in the $V$-band using a 28~cm telescope and modelled using the {\sc ebop} code \cite{Etzel81conf,PopperEtzel81aj}. They found night-to-night magnitude offsets in these data, concluded that they were likely of instrumental origin, and removed them before analysis. They determined the physical properties of the component stars and found them to be consistent with theoretical predictions for a solar metallicity and a young age (40~Myr but with a large uncertainty). Finally, they suspected $\delta$~Scuti pulsations in star~A due to the night-to-night offsets and its much larger scatter in the RVs compared to those for star~B.

\targ\ has been mentioned in several other works, and additional times of mid-eclipse have been published. However, L06 remains the only detailed analysis of the system prior to the current work.


\begin{figure}[t] \centering \includegraphics[width=\textwidth]{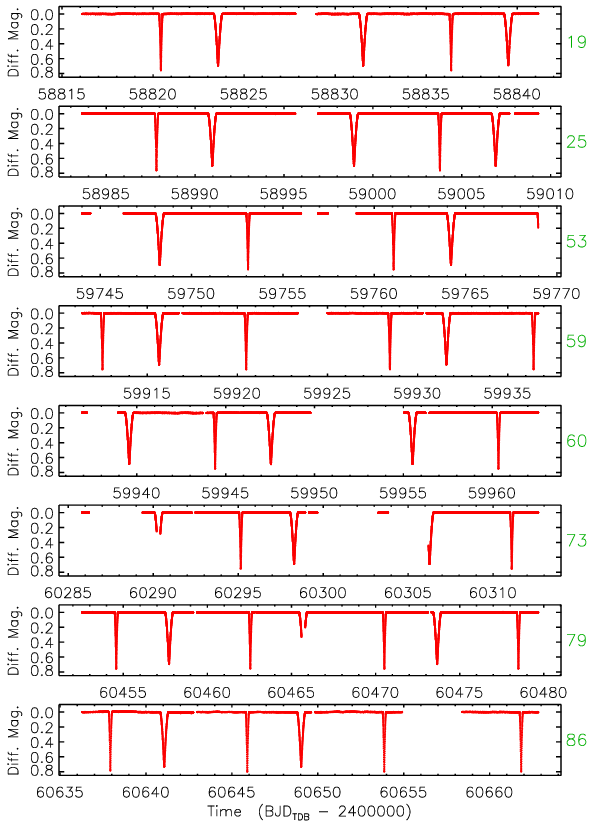} \\
\caption{\label{fig:time} Light-curves of \targ\ from all eight TESS sectors. The flux 
measurements have been converted to magnitude units and the median subtracted.} \end{figure}

\clearpage

\section*{Photometric observations}


\targ\ was observed by the Transiting Exoplanet Survey Satellite (TESS \cite{Ricker+15jatis}) mission during eight sectors. Light-curves from the TESS Science Processing Center (SPOC \cite{Jenkins+16spie}) at a cadence of 120~s are available for seven of the sectors (19, 25, 53, 59, 60, 73 and 79). Quick Look Pipeline ({\sc qlp} \cite{Huang+20rnaas}) data at 200~s cadence are available for the last sector (86).

We downloaded these data from the Mikulski Archive for Space Telescopes (MAST\footnote{\texttt{https://mast.stsci.edu/portal/Mashup/Clients/Mast/Portal.html}}) using the {\sc lightkurve} package \cite{Lightkurve18}, with the ``hard'' flag to reject data of lower quality. The light-curve from each sector was then converted into differential magnitude and the median magnitude was subtracted to normalise the data to zero outside eclipse. The light-curves are shown in Fig.~\ref{fig:time}.

We queried the \gaia\ DR3 database\footnote{\texttt{https://vizier.cds.unistra.fr/viz-bin/VizieR-3?-source=I/355/gaiadr3}} for sources within 2~arcmin of \targ, finding 41 including the dEB itself. The brightest of these is 5.3~mag fainter than \targ\ in the \Grp\ band, suggesting that contamination of the TESS light-curve by nearby stars is negligible.


\section*{Light-curve analysis}

The TESS light-curves for each sector all have full coverage of at least two primary and two secondary eclipses, with the exception of sector 73 for which there are two primary but only one secondary eclipse. We therefore modelled the light-curves from each sector independently. Prior to this, we extracted the data for each fully-observed eclipse from the TESS light-curves in order to reject data which were not useful but would still incur computational overhead. We included an additional 0.2~d of data before the start and after the end of each eclipse, to which we fitted straight line. This fit was then subtracted from the data to normalise each eclipse to zero differential magnitude.

\begin{figure}[t] \centering \includegraphics[width=\textwidth]{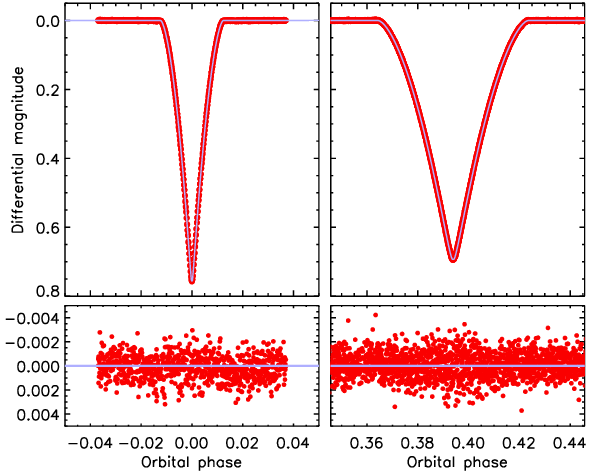} \\
\caption{\label{fig:phase} {\sc jktebop} best fit to the light-curve of \targ\ from 
TESS sector 19 for the primary eclipse (left panels) and secondary eclipse (right 
panels). The data are shown as filled red circles and the best fit as a light blue 
solid line. The residuals are shown on an enlarged scale in the lower panels.} \end{figure}

The light-curve for each sector was then modelled using version 45 of the {\sc jktebop}\footnote{\texttt{http://www.astro.keele.ac.uk/jkt/codes/jktebop.html}} code \cite{Me++04mn2,Me13aa}. We fitted for the orbital period ($P$), a reference time of primary minimum ($T_0$) selected as the primary eclipse closest to the midpoint of the sector, the sum and ratio of the fractional radii ($r_{\rm A}+r_{\rm B}$ and $k = r_{\rm B}/r_{\rm A}$), the central surface brightness ratio ($J$), third light ($L_3$), and orbital inclination ($i$). We included orbital eccentricity ($e$) by fitting for $e\cos\omega$ and $e\sin\omega$, where $\omega$ is the argument of periastron.

Limb darkening (LD) was incorporated using the power-2 law \cite{Hestroffer97aa,Maxted18aa,Me23obs2}. The two stars have very similar properties so we used the same LD coefficients for both. We fitted for the linear coefficient ($c$) and fixed the nonlinear coefficient ($\alpha$) at a theoretical value \cite{ClaretSouthworth22aa,ClaretSouthworth23aa}. The best fit for the data from sector 19 is shown in Fig.~\ref{fig:phase}; other sectors are similar. The fitted parameters are given in Table~\ref{tab:jktebop} and are the unweighted mean of the individual values for each parameter.

\begin{table} \centering
\caption{\em \label{tab:jktebop} Photometric parameters of \targ\ measured using 
{\sc jktebop} from the TESS light-curves. The error bars are 1$\sigma$ and are the 
largest of three options calculated for each parameter. Values from L06 (their table~6) 
are shown for comparison; the two bracketed quantities came instead from their table~5.}
\begin{tabular}{l r@{\,$\pm$\,}l r@{\,$\pm$\,}l}
{\em Parameter}                           & \multicolumn{2}{c}{\em This work} & \multicolumn{2}{c}{\rm L06} \\[3pt]
{\it Fitted parameters:} \\
Orbital inclination (\degr)               &   89.903   & 0.013   &   89.89    & 0.03     \\
Sum of the fractional radii               &    0.12102 & 0.00008 & \multicolumn{2}{c}{}  \\
Ratio of the radii                        &    0.9701  & 0.0017  &    1.005   & 0.012    \\
Central surface brightness ratio          &    0.9577  & 0.0010  &    0.931   & 0.005    \\
Third light                               & $-$0.0001  & 0.0149  & \multicolumn{2}{c}{}  \\
$e\cos\omega$                             & $-$0.15291 & 0.00028 & $-$0.14897 & 0.00013  \\
$e\sin\omega$                             &    0.41275 & 0.00037 &    0.4224  & 0.0017   \\
LD coefficient $c$                        &    0.597   & 0.014   & \multicolumn{2}{c}{}  \\
LD coefficient $\alpha$                   & \multicolumn{2}{c}{0.4555 (fixed)}           \\
{\it Derived parameters:} \\
Fractional radius of star~A               &    0.06143 & 0.00006 &   0.0603   & 0.0004   \\
Fractional radius of star~B               &    0.05959 & 0.00008 & \multicolumn{2}{c}{}  \\ 
Light ratio $\ell_{\rm B}/\ell_{\rm A}$   &    0.9013  & 0.0028  &   0.961    & 0.026    \\
Orbital eccentricity                      &    0.44017 & 0.00035 & (\,0.4415   & 0.012\,) \\
Argument of periastron ($^\circ$)         &  110.328   & 0.041   & (\,109.78  & 0.07\,)  \\
\end{tabular}
\end{table}

We determined three sets of uncertainties for the fitted and derived parameters. The first set is the r.m.s.\ scatter of the fitted values from the different sectors, the second set was obtained by running 1000 Monte Carlo simulations for each sector, and the third set was from running residual-permutation simulations for each sector \cite{Me08mn}. We took the largest of the three options for each parameter, which in most cases was from the scatter of the fitted values. These are also given in Table~\ref{tab:jktebop}. We did not divide by $\sqrt{8}$ to obtain the standard error, because this led to error bars which are smaller than the limits to which our modelling code has been proven \cite{Maxted+20mn}.

\begin{figure}[t] \centering \includegraphics[width=\textwidth]{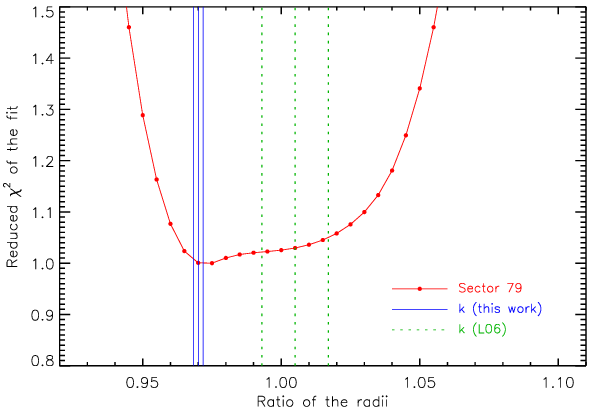} \\
\caption{\label{fig:kfix} Quality of the fit, \chir, to the TESS sector 79 light 
curve of \targ\ as a function of ratio of the radii, $k$ (red lines and filled 
circles). The blue solid lines indicate the value and its uncertainty from the 
current work (based on all eight TESS sectors, not just sector 79), and the green 
dotted lines show the best value and uncertainty found by L06.} \end{figure}

The parameter values we measured are significantly different to those from L06. These are collected in Table~\ref{tab:jktebop} for convenience. The most obvious discrepancies are for $k$ and the light ratio, which of course are physically related. We investigated this by running a set of {\sc jktebop} fits to the data from TESS sector 79, with $k$ fixed at a grid of values between 0.6 and 1.5. The quality of the fit is shown in Fig.~\ref{fig:kfix}, together with the $k$ found in this work and by L06 (Table~\ref{tab:jktebop}). It can be seen that our $k$ measurement is much more precise and also significantly different to that of L06, driven primarily by the much better quality of the TESS data. 

The light ratio we find ($0.901 \pm 0.003$) does not match either of the spectroscopic light ratios measured by L06 ($1.06 \pm 0.06$ from their KPNO spectra and $1.32 \pm 0.02$ from their FLWO spectra). It is also moderately different to L06's photometric value ($0.96 \pm 0.03$), which can be attributed to the different $k$ we obtain as discussed above. The two spectroscopic light ratios are inconsistent with each other and with the photometric values, as noted by L06. A plausible explanation would be that one component is a metallic-lined star, thus the ratio of spectral line strengths differs from the continuum light ratio, but this is ruled out by L06's observation that the two stars are chemically normal. The discrepancy remains unexplained.


\section*{Orbital ephemeris}

\begin{table} \centering
\caption{\em Times of mid-eclipse for \targ\ and their residuals versus the fitted ephemeris. \label{tab:tmin}}
\setlength{\tabcolsep}{10pt}
\begin{tabular}{rrrrr}
{\em Orbital} & {\em Eclipse time}  & {\em Uncertainty} & {\em Residual} & {\em TESS}   \\
{\em cycle}   & {\em (BJD$_{TDB}$)} & {\em (d)}         & {\em (d)}      & {\em sector} \\[3pt]

$-139.0$ & 2458836.367724 & 0.000012 & $ 0.000057$ & 19 \\   
$-118.0$ & 2459003.768820 & 0.000014 & $-0.000003$ & 25 \\   
$ -23.0$ & 2459761.059727 & 0.000012 & $-0.000039$ & 53 \\   
$  -2.0$ & 2459928.460878 & 0.000007 & $-0.000043$ & 59 \\   
$   0.0$ & 2459944.403874 & 0.000012 & $-0.000015$ & 60 \\   
$  44.0$ & 2460295.149249 & 0.000012 & $ 0.000082$ & 73 \\   
$  66.0$ & 2460470.521820 & 0.000007 & $ 0.000013$ & 79 \\   
$  89.0$ & 2460653.865933 & 0.000010 & $ 0.000003$ & 86 \\   
\end{tabular}
\end{table}

We determined a new orbital ephemeris for \targ\ using the times of effective primary eclipse found in the fits to the individual TESS sectors above. A straight line fitted to these data gives the ephemeris
\begin{equation}
\mbox{Min~I} = {\rm BJD}_{\rm TDB}~ 2459944.403889 (14) + 7.97148361 (20) E
\end{equation}
where $E$ is the number of cycles since the reference time of minimum and the bracketed quantities indicate the uncertainty in the final digit of the previous number. The times of minimum and residuals of the fit are reported in Table~\ref{tab:tmin}.

\begin{figure}[t] \centering \includegraphics[width=\textwidth]{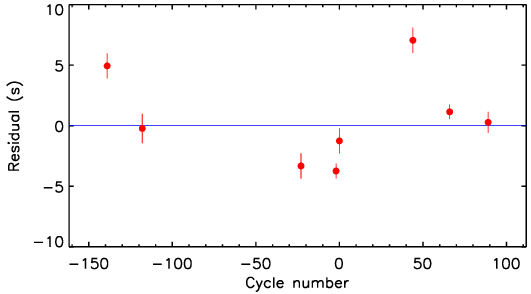} \\
\caption{\label{fig:tmin} Residuals of the times of minimum light from Table~\ref{tab:tmin} 
versus the best-fitting ephemeris. Red open circles indicate the values calculated from 
TESS data in the current work. The blue solid line indicates a residual of zero.} \end{figure}

The residuals of the fit are plotted in Fig.~\ref{fig:tmin} and show a scatter which is small in absolute terms (the r.m.s.\ is 3.6~s) but bigger than the error bars. The reduced $\chi^2$ of the fit is 15.3 and there are hints of a periodic variation which could be explained by the light-time effect due to a third body on an eccentric orbit. The uncertainties in the ephemeris above have been increased by $\sqrt{15.3}$ to account for this excess scatter. Further measurements are required to investigate the possible light-time effect.

We tested including eclipse timings from the literature to see if these might support the third-body hypothesis. The measurement errors from the seven individual timings given by L06 are too large to constrain the solution, given the high quality of the TESS data. The time system used for these timings is also not clear: they are labelled as HJD but without indication of whether they are UTC or TDB. The difference between these two time systems is much larger than the size of the prospective light-time effect. Five more timings exist in the literature \cite{Hubscher++05ibvs,HubscherWalter07ibvs,Hubscher+10ibvs,Hubscher11ibvs,Hubscher15ibvs} but all are for secondary eclipses and have a larger measurement error than the L06 timings. We thus leave confirmation or refutation of this effect to future work.


\section*{Radial velocity analysis}

\begin{figure}[t] \centering \includegraphics[width=\textwidth]{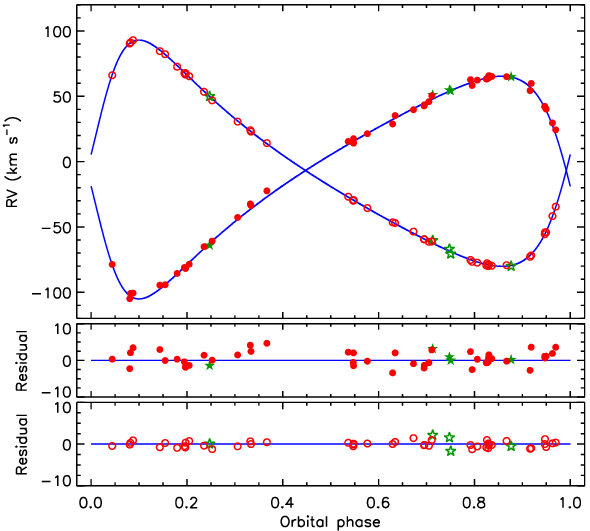} \\
\caption{\label{fig:rv} RVs of \targ\ from L06 compared to the best fit from 
{\sc jktebop} (solid blue lines). The RVs for star~A are shown with filled 
symbols, and for star~B with open symbols. The RVs from the FLWO spectra are
shown as red circles, and from the KPNO spectra as green stars. The residuals 
are given in the lower panels separately for the two components.} \end{figure}

L06 tabulated RVs for both components of \targ\ from their five KPNO spectra and 47 FLWO spectra. We reanalysed the RVs using {\sc jktebop}, fixing the values of $P$, $T_0$, $e\cos\omega$ and $e\sin\omega$ to those in Table~\ref{tab:jktebop}. We fitted for the velocity amplitudes ($K_{\rm A}$ and $K_{\rm B}$) and the systemic velocities ($V_{\rm \gamma,A}$ and $V_{\rm \gamma,B}$) of the two stars. We also allowed for an orbital phase offset, which in all cases was very small and consistent with the difference between the TDB timescale used in the current work and the UTC timescale probably used by L06. We calculated orbits based on only the FLWO RVs, and on all RVs assuming the two sources are on a consistent velocity scale. The results are given in Table~\ref{tab:sb} and a representative orbit is plotted in Fig.~\ref{fig:rv}

\begin{sidewaystable} \centering
\caption{\em Spectroscopic orbits for \targ\ from the literature and from the current work. 
All quantities are given in km~s$^{-1}$. We give both sets of orbits from L06: (a) based on 
all (KPNO and FLWO) RVs; (b) based on all RVs and times of minimum light. The orbits from 
the current work were calculated using either the FLWO RVs or all RVs. \label{tab:sb}}
\setlength{\tabcolsep}{10pt}
\begin{tabular}{lccccccc}
{\em Source} & {\em $K_{\rm A}$} & {\em $K_{\rm B}$} & {\em $V_{\rm \gamma}$} & {\em $V_{\rm \gamma,A}$} & {\em $V_{\rm \gamma,B}$} & {\em $\sigma_{\rm A}$} & {\em $\sigma_{\rm B}$} \\[10pt]
L06 (a)                 & $85.27 \pm 0.41$ & $86.64 \pm 0.15$ & $-6.79 \pm 0.10$ &                  &                  & 2.07 & 0.64 \\
L06 (b)                 & $85.23 \pm 0.39$ & $86.62 \pm 0.14$ & $-6.77 \pm 0.10$ &                  &                  &      &      \\[5pt]
This work (FLWO)        & $84.94 \pm 0.39$ & $86.57 \pm 0.13$ &                  & $-6.28 \pm 0.28$ & $-6.90 \pm 0.09$ & 1.91 & 0.65 \\
This work (FLWO)        & $85.06 \pm 0.42$ & $86.58 \pm 0.14$ & $-6.84 \pm 0.09$ &                  &                  & 2.00 & 0.65 \\[5pt]
This work (KPNO + FLWO) & $85.07 \pm 0.38$ & $86.56 \pm 0.13$ &                  & $-6.27 \pm 0.27$ & $-6.89 \pm 0.10$ & 1.88 & 0.76 \\
This work (KPNO + FLWO) & $85.22 \pm 0.38$ & $86.58 \pm 0.13$ & $-6.83 \pm 0.09$ &                  &                  & 1.96 & 0.76 \\
\end{tabular}
\end{sidewaystable} 

All spectroscopic orbits in Table~\ref{tab:sb} agree within their uncertainties, making it unimportant which is adopted for the rest of the analysis. We chose to use the orbits based on only the FLWO data -- to avoid exposure to any velocity offsets between the two spectrographs -- and with separate systemic velocities for the two stars. Our results confirm the observation of L06 that the scatter of the RVs for star~A is much larger than that for star~B.

\section*{Physical properties and distance to \targ}

We calculated the physical properties of \targ\ using our usual method: the {\sc jktabsdim} code \cite{Me++05aa} fed with the results of the photometric and spectroscopic analyses above. The properties are collected in Table~\ref{tab:absdim}. The masses of the two stars are measured to 0.5\% and 0.9\% precision, respectively, and agree very well with the values from L06 (as expected). The radii are measured to 0.3\% precision and do not agree with those from L06, because we found a smaller $k$ (as discussed above). 

L06 found a mean effective temperature of the system of $7030 \pm 150$~K. We determined individual temperatures from this, the radius and surface brightness ratios, and the equations derived in our study of V454~Aur \cite{Me24obs3}. The values are also given in Table~\ref{tab:jktebop}. 

\begin{table} \centering
\caption{\em Physical properties of \targ\ defined using the nominal solar units 
given by IAU 2015 Resolution B3 (ref.~\citenum{Prsa+16aj}). \label{tab:absdim}}
\begin{tabular}{lr@{~$\pm$~}lr@{~$\pm$~}l}
{\em Parameter}        & \multicolumn{2}{c}{\em Star A} & \multicolumn{2}{c}{\em Star B}     \\[3pt]
Mass ratio   $M_{\rm B}/M_{\rm A}$                  &  \multicolumn{4}{c}{$0.9811 \pm 0.0047$}        \\
Semimajor axis of relative orbit (\Rsunnom)         &  \multicolumn{4}{c}{$24.265 \pm 0.058$}         \\
Mass (\Msunnom)                                     &   1.523  & 0.008       &   1.494  & 0.014       \\
Radius (\Rsunnom)                                   &   1.4906 & 0.0039      &   1.4460 & 0.0040      \\
Surface gravity ($\log$[cgs])                       &   4.2740 & 0.0011      &   4.2921 & 0.0023      \\
Density ($\!\!$\rhosun)                             &   0.4600 & 0.0017      &   0.4942 & 0.0023      \\
Synchronous rotational velocity ($\!\!$\kms)        &   9.461  & 0.025       &   9.177  & 0.025       \\
Pseudo-synchronous rotational velocity ($\!\!$\kms) &  27.104  & 0.073       &  26.293  & 0.076       \\
Effective temperature (K)                           &   7070   & 150         &   6990   & 150         \\
Luminosity $\log(L/\Lsunnom)$                       &   0.699  & 0.037       &   0.653  & 0.037       \\
$M_{\rm bol}$ (mag)                                 &   2.99   & 0.09        &   3.11   & 0.09        \\
Interstellar reddening \EBV\ (mag)                  &  \multicolumn{4}{c}{$0.025 \pm 0.015$}          \\
Distance (pc)                                       &  \multicolumn{4}{c}{$300.3 \pm 3.8$}            \\[3pt]       
\end{tabular}
\end{table}


Table~\ref{tab:absdim} also reports the synchronous and pseudo-synchronous rotational velocities of the stars. The former is the rotational velocity giving one stellar rotation per orbital period. The latter is the rotational velocity resulting in synchronous rotation at periastron, calculated using eq.~44 from Hut \cite{Hut81aa}. Because tidal effects are strongest at periastron (as this is when the stars are closest), it is expected that the stars are rotating pseudo-synchronously. However, their measured rotational velocities are consistent with synchronous rather than pseudo-synchronous rotation. The reason for this is unclear. The age of the system (see below) puts it much older than the rotational synchronisation timescale (40~Myr) but much younger than the orbital circularisation timescale (20~Gyr) according to the theory of Zahn \cite{Zahn77aa}.

We determined the distance to the system using the $UBV$ magnitudes from Lacy \cite{Lacy92aj}, the $JHK_s$ magnitudes from 2MASS \cite{Cutri+03book} transformed onto the Johnson system, and the bolometric corrections from Girardi et al.\ \cite{Girardi+02aa}. A small interstellar reddening of $\EBV = 0.025 \pm 0.015$ was needed to bring the optical distances into agreement with the infrared ones. Our final distance measurement is from the $K$-band and is $300.3 \pm 3.8$~pc, which is in excellent agreement with the distance of $301.2 \pm 1.2$~pc from inverting the \gaia\ DR3 parallax.

We compared the masses, radii and temperatures of the components of \targ\ to the predictions of the {\sc parsec} 1.2 evolutionary models \cite{Bressan+12mn}. The mass-radius relation defined by the two stars is steeper than that predicted by the models, but agreement within 1$\sigma$ can be obtained for a metal abundance fraction by mass of $Z=0.020$ and an age of 220~Myr. The predicted temperatures for this age and $Z$ are approximately 50~K higher than observed, so are well within the error bars. Our results here differ from those of L06, who required the system to be on or around the zero-age main sequence in order to explain their measurement of a larger radius and lower surface gravity for star~B versus star~A.

\section*{Summary and conclusions}

\targ\ is a dEB containing two F0~V stars in an orbit of period 7.97~d and eccentricity 0.44. We have determined the physical properties of the component stars using light-curves from eight TESS sectors and the RVs from L06. The masses are measured to 1\% and the radii to 0.3\%. We find the less massive star to have a smaller radius than its companion, as expected from stellar evolution theory, resolving a discrepancy found in previous work. Our measured distance to the system agrees well with that from the \gaia\ DR3 parallax. The properties of the stars are consistent with theoretical predictions for a slightly super-solar metallicity and an age of 220~Myr. There is a hint of periodic variation in the times of minimum light suggestive of the light-time effect from a third component on an eccentric orbit.

Our original interest in this object was piqued by the possibility of pulsations in its light-curve. We investigated this by fitting the most suitable dataset, TESS sectors 59 and 60, with {\sc jktebop} and calculating a periodogram of the residuals. The highest frequency in the amplitude spectrum was at 0.546\cd\ and had an amplitude of 0.26~mmag and a signal-to-noise ratio of 4.4. This signal-to-noise ratio is below that required for the detection be significant \cite{BaranKoen21aca}. We therefore analysed the data from sectors 79 and 86 as well to see if they could confirm the detection. We did not find the signal present in either sector. We conclude that pulsations are not detectable in the currently available data of \targ.


\section*{Acknowledgements}
 
We thank the anonymous referee for a helpful report.
We acknowledge support from STFC under grant number ST/Y002563/1.
This paper includes data collected by the TESS\ mission and obtained from the MAST data archive at the Space Telescope Science Institute (STScI). Funding for the TESS mission is provided by the NASA's Science Mission Directorate. STScI is operated by the Association of Universities for Research in Astronomy, Inc., under NASA contract NAS 5–26555.
This work has made use of data from the European Space Agency (ESA) mission {\it Gaia}\footnote{\texttt{https://www.cosmos.esa.int/gaia}}, processed by the {\it Gaia} Data Processing and Analysis Consortium (DPAC\footnote{\texttt{https://www.cosmos.esa.int/web/gaia/dpac/consortium}}). Funding for the DPAC has been provided by national institutions, in particular the institutions participating in the {\it Gaia} Multilateral Agreement.
The following resources were used in the course of this work: the NASA Astrophysics Data System; the SIMBAD database operated at CDS, Strasbourg, France; and the ar$\chi$iv scientific paper preprint service operated by Cornell University.



\end{document}